\documentclass[prl,aps,twocolumn,showpacs,showkeys]{revtex4}
\usepackage{epsf,amssymb,amsmath}

\begin{document}
\title{ \sffamily\bfseries\Large
Evolution of the Protein Interaction Network of Budding Yeast: \\
Role of the Protein Family Compatibility Constraint\\}

\author{\sc K.-I. Goh, B. Kahng, and D. Kim}

\affiliation{\mbox{School of Physics and Center for 
Theoretical Physics, Seoul National University NS50, 
Seoul 151-747, Korea}}
\date{\today}

\begin{abstract}
Understanding of how protein interaction networks (PIN) of living 
organisms have evolved or are organized can be the first stepping 
stone in unveiling how life works on a fundamental basis.
Here, we introduce a new {\em in-silico} evolution model of
the PIN of budding yeast, {\em Saccharomyces cerevisiae}; 
the model is composed of the PIN and the protein family network. 
The basic ingredient of the 
model includes family compatibility which constrains
the potential binding ability of a protein,
as well as the previously proposed 
gene duplication, divergence, and mutation.
We investigate various structural properties of our model network 
with parameter values relevant to budding yeast and 
find that the model successfully reproduces the 
empirical data. 
\end{abstract}
\maketitle

Studying complex systems by means of their network representation
has attracted much attention recently \cite{rmp,advphys,siam,saemulli,dslee,han}.
The cell, one of the best examples of complex systems, can also
be viewed as a network:
The cellular components, such as genes, proteins, and other 
biological molecules, connected by all physiologically
relevant interactions, form a full weblike molecular architecture
in a cell~\cite{pyramid,network-biology}. 
Among the various levels, the protein interaction network (PIN) 
plays a pivotal role as it acts as a basic physical protocol 
of cooperative functioning in many physiological processes.
In the PIN, proteins are viewed 
as nodes, and two proteins are linked if they physically 
contact each other. 
Thanks to recent progress in high-throughput experimental techniques, 
the data set of protein interactions for budding yeast,
{\em Saccharomyces cerevisiae}, has been firmly 
established in the last few years \cite{uetz,ito,gavin,ho,tong,mips,dip,bind}. 
Thus, it offers a good testbed to understand how it has evolved
to form its status quo from basic evolutionary rules.
In this paper, our aim is to introduce a simple evolutionary model
to reproduce the structural properties of the PIN of budding yeast,
thereby deepening our understanding of the driving force for
cellular evolution.

At a certain level of abstraction, one may view a protein as 
an assembly of domains. It is domains that offer structural 
and functional units. They act as basic units in 
the interactions between proteins and in the evolution 
of protein structures. Proteins are grouped into so-called protein 
families or superfamilies
according to the domain structure within them \cite{alberts}.
The proteins within a family are monophyletic;
that is, they originate from a common ancestor
and are fairly well conserved during evolution. 
The protein family network (PFN) is defined as the one 
whose nodes are protein families, and two families are connected 
if any of the domains within them simultaneously
occur in a single protein or any proteins within
them interact with each other \cite{jpark}.
The distributions of the degrees and the sizes of the families in the PFN
also follow power laws \cite{jpark,huynen}.
Given that the entities of proteins and protein families
are not separable but linked via domains as intermediates,
it is desirable to unify their evolutions into a single framework. 

So far, several {\it in-silico} evolution models have been proposed 
for the yeast PIN \cite{sole,vazquez,berg,kim,chung}.
A distinguishing aspect in the evolution of the PIN compared
with that of other complex networks is the concept of ``evolution 
by duplication''~\cite{ohno}:
A new protein is thought to be created mainly by gene duplication.
Subsequently, the duplicate protein may lose redundant interactions
endowed from its ancestor to reduce redundancy, 
which is called divergence (or diversification).
A protein also gains new interactions with other
proteins via mutation. These three processes,
duplication--divergence--mutation, have been regarded as the basic
ingredients in the evolution of the PIN. While those {\it in-silico}
models~\cite{sole,vazquez,kim,chung,berg}
were successful in generating a fat-tail or power-law behavior in
the degree distribution,  
they hardly reproduced other structural properties of the yeast 
PIN network, such as the clustering coefficient, the assortativity,
{\it etc.}, which we will specify in more detail shortly. 
The model we introduce here, however, can incorporate other 
structural properties of the yeast PIN as well as the degree distribution.
To this end, we introduce the concept of 
``family compatibility'' (FC):
An interaction between two proteins is possible only when
the corresponding families they belong to are compatible,
and only those families linked via the PFN are compatible with one another.
With this, we realize the effective structural constraint 
in physical binding between proteins, which is coupled with
the evolutionary lineage of proteins through the notion of protein family.

\begin{figure}[t]
\centerline{\epsfxsize=9cm \epsfbox{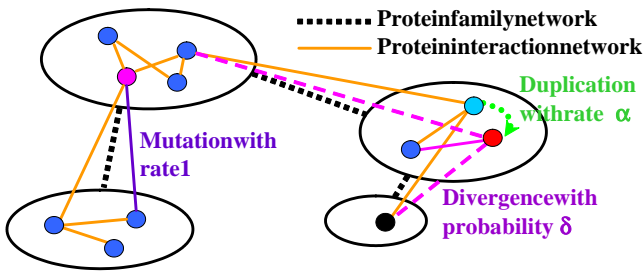}}
\caption{Schematic picture of the evolution rule of the model.
The elementary steps are composed of i) duplication 
(light blue protein $\rightarrow$ red protein), 
ii) divergence (dashed pink links), and 
iii) mutation (violet link from the pink protein).
In addition, the mutation is constrained by family 
compatibility; for example, the pink protein cannot 
interact with the black protein because they are not compatible.
}
\end{figure}

\begin{figure*}
\centerline{\epsfxsize=15cm \epsfbox{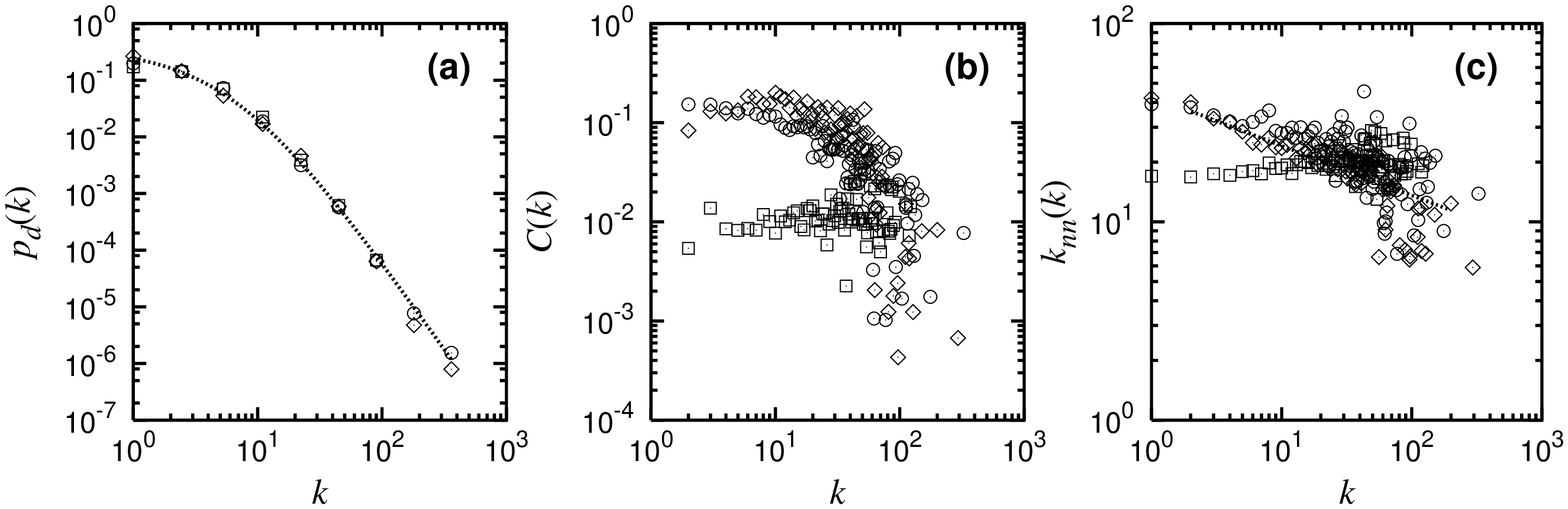}}
\caption{
Simulation results ($\bigcirc$) of the model agree well with the 
empirical data ($\diamond$).
Shown are 
(a) the degree distribution $P(k)$,
(b) the hierarchical clustering $C(k)$, and
(c) the average neighbor-degree function 
$\langle k_{\rm nn}\rangle$ for the protein interaction network.
The dotted line in (a) is a fit to Eq.~(\ref{pk}).
The results of the model without FC ($\Box$), which fail
to reproduce the empirical features, are also shown for 
comparison.
}
\end{figure*}

{\em Model}--- The model can be depicted schematically as in Fig.~1.
The whole system is composed of two types of networks, 
the PIN and the PFN. A number of proteins are grouped, forming 
a protein family. Protein families link to other protein families,
forming the PFN.
Two proteins belonging to different protein families can 
interact only when the respective families are also linked.
Each family has a fitness-like parameter, the number of domains
within it, $D_f$, which is not fixed, but evolves with the PFN.
The evolution takes place in two stages. In the first stage,
the protein families are created along with the proteins;
thus, the PFN coevolves with the PIN.
In the second stage, the PFN is kept fixed, and the evolution of 
the PIN continues on top of it. 
A detailed description of the procedure is as follows:

\begin{enumerate}
\item Initially, there are $n_0$ proteins, each of which constitutes
its own protein family. All $n_0$ proteins
are interconnected with one another, as are the $n_0$ protein families.
We choose $n_0=3$ to be minimal. 
Each family has $D_f=2$ domains, the number of family-links it has.

\item In the first stage, proteins and protein families coevolve: 
At each step, with rate $\alpha$, a new protein, say $a$, is created 
by duplicating an existing protein $b$ chosen randomly. The new protein $a$
creates its own protein family $F_a$. 
Each of the inherited interactions of the protein $a$
is removed with probability $\delta$, a process called divergence.
Through divergence, the degree of the new protein $a$, $k_a$,
usually becomes less than that of the mother protein $k_b$.
The linkage of the new protein family is determined by that of 
the protein created. By this process, the newly born family $F_a$ 
consists of a single protein, but has a number of linkages, say $K_{F_a}$, 
to existing families. 
The initial number of domains in the family is set to 
$D_{F_a}=K_{F_a}$. In some cases, the newly created protein is left with no 
interaction at all $(K_{F_a}=0)$. 
In this case, we do not let it establish a new 
family, but regard it as a remnant in the previous family.
When this case happens, the population of the family to which 
the duplicated protein belongs is increased by 1. Note that the 
remnant can later gain new interactions via mutation described below
and join the protein interaction network.

With rate $1$, a randomly chosen existing protein $i$ gains a new 
interaction to another previously unlinked protein $j$, which is 
chosen among the proteins within compatible families,
according to the probability,
\begin{equation}
\Pi_j= \dfrac{D_{F_j}}{\underset{F_{l}\leftrightarrow F_i}{\displaystyle \sum\nolimits} D_{F_{l}}},
\end{equation}
where $F_i$ means the family
to which the protein $i$ belongs and $X\leftrightarrow Y$ means that 
the families $X$ and $Y$ are compatible, i.e., linked in the PFN.
Eq.~(1), the preferential attachment in the domain abundance
constrained by FC, makes our model distinct
and successful.
In this process, the mutation as we will call it, the number of domains 
in the family $F_i$ increases by 1, but the number of domains in $F_j$
does not. 
This accounts for the acquisition of a new domain via mutation in 
the family $F_i$. This stage lasts until there are 1,000 proteins 
made, during which about $500$$\sim$$600$ families are created, a number 
comparable with the number of superfamilies in yeast~\cite{superfamily}

\item 
In the second stage, the same protein evolution process as in 
the first stage occurs, except that the PFN is 
kept fixed and the daughter protein remains in the same family as 
its mother in the duplication process.
This stage lasts until there are about 6,000 proteins in
the network, the approximate size of the yeast proteome. 
\end{enumerate}

A few remarks on the model are in order.
First, this model is designed to be as simple as possible while 
implementing FC into the 
trio of duplication, divergence, and mutation,
which we believe to be the most basic processes.
Many interesting processes, such as lateral gene transfer
and {\it de novo} creation of proteins and protein families,
are not covered in this model, however.
Second, we made an assumption that the time-scale of
the PFN evolution is strictly separated,
which might be an oversimplification.
Third, proteins and protein families may become extinct during evolution,
followed by the loss of the interactions between them.
However, we may view the parameters of the evolution rates,
such as $\alpha$ and $\delta$,
as {\it effective} ones incorporating all these details. 
Also, for the sake of minimizing the number of free parameters,
we assume that the duplication and the divergence rates of proteins 
and protein families are equal, i.e., $\alpha=\alpha_f$ and 
$\delta=\delta_f$, although we can fix $\alpha$ and $\delta$ for any 
given set of ($\alpha_f$, $\delta_f$) to incorporate the empirical 
data. 

{\em Structure of the yeast PIN}--- 
Several analyses on the topological properties of the yeast
PIN have been performed during recent
years \cite{lethal,maslov,wagner}. Since then, however, new 
protein--protein interactions in yeast have been discovered steadily,
so we repeat the analysis by integrating the most up-to-date data 
from various public resources, such as
(i) the database at the Munich Information Center for Protein Sequences \cite{mips}, 
(ii) the database of the interacting proteins \cite{dip}, 
(iii) the biomolecular interaction network database \cite{bind},
(iv) the two-hybrid datasets obtained by Uetz {\it et al.}~\cite{uetz}, 
by Ito {\it et al.}~\cite{ito}, and by Tong {\it et al.}~\cite{tong},
and (v) the mass spectrometry data (filtered) by Ho {\it et al.}~\cite{ho}.
After trimming the synonyms and other redundant entries manually,
the resulting network consists of 15,\mbox{ }652 interactions
(excluding self-interactions) between 4,\mbox{ }926 nodes (in terms of
distinct open reading frames and other biomolecules).

The topological properties of the integrated yeast PIN are shown 
in Fig.~2:

(a) The degree distribution of the PIN fits well to the generalized Pareto 
distribution (or a generalized power law) \cite{ab,koonin},
\begin{equation}
p_d(k)\sim (k+k_0)^{-\gamma},
\label{pk}
\end{equation}
with $k_0=8.0$ and $\gamma\simeq3.45$.
Note that different functional types of the degree distribution from 
Eq.~(\ref{pk}) were proposed~\cite{sole,vazquez,berg,wagner,lethal} 
based on smaller-scale datasets than the current one.  

(b) The yeast PIN is highly clustered and modular. 
To quantify this, we measured the local clustering of a protein $i$,
$c_i = {2e_i}/{k_i(k_i-1)}$, where $e_i$ is the number of links 
present between the $k_i$ neighbors of node $i$ out of its maximum 
possible number $k_i(k_i-1)/2$.
The clustering coefficient of a graph, $C$, is the average of 
$c_i$ over all nodes with $k_i\ge 2$. We obtain $C\approx 0.128$. 
$C(k)$ is the clustering function of vertices with degree 
$k$~\cite{vespignani2,ravasz}.
$C(k)$ exhibits a plateau for small $k$ while it drops rapidly 
for large $k$.
Such a plateau in the clustering function may reflect the 
functional module structure within the PIN, inside which the 
network is denser due to the high cooperativity to perform
a given cellular task. Such locally dense modules are interconnected
by a few global mediators, which are likely to be the hubs in the PIN \cite{han-vidal}.
This feature is what most existing PIN models fail to reproduce.
As we will show, the FC constraint that we introduce 
successfully accounts for the emergence of the plateau in $C(k)$.

(c) The yeast PIN shows a dissortative degree correlation.
The average neighbor-degree function 
$\langle k_{\rm nn}\rangle(k)$ \cite{knn} is measured to be
$\langle k_{\rm nn} \rangle(k) \sim k^{-\nu}$
with $\nu\approx 0.3$, somewhat smaller than the value reported based 
on a single two-hybrid dataset alone~\cite{maslov}.
The assortativity $r$, defined as the Pearson correlation coefficient 
between the degrees of the two vertices on each side of 
a link~\cite{assort}, is measured to be $r \approx -0.13$.
In Table \ref{tab1}, we summarize our measurements for the topological properties 
of the integrated yeast PIN. 
\begin{table}[b]
\caption{Topological quantities of the integrated 
yeast PIN and the model network. 
Error bars in the model results are the standard deviations of the
quantities from 1000 runs.}
\label{tab1}
\begin{ruledtabular}
\begin{tabular}{lll}
item & model & yeast PIN \\
\hline
total number of nodes $n$\phantom{aaa} & 6000\phantom{aaa} & $\approx$6000 \\
number of interacting nodes $N$\phantom{aaa} & 5079$\pm$54 & 4926 \\
average degree $\langle k\rangle$\phantom{aaa} & 6.5$\pm$0.3 & 6.35 \\
clustering coefficient $C$ & 0.13$\pm$0.02 & 0.128 \\
assortativity index $r$ & $-$0.09$\pm$0.04 & $-0.13$ \\
size of the largest component $N_1$ & 5051$\pm$53 & 4832 \\
\end{tabular}
\end{ruledtabular}
\end{table}

\begin{figure*}
\begin{minipage}[!t]{0.5\linewidth}
\flushright{\epsfxsize=6.3cm \epsfbox{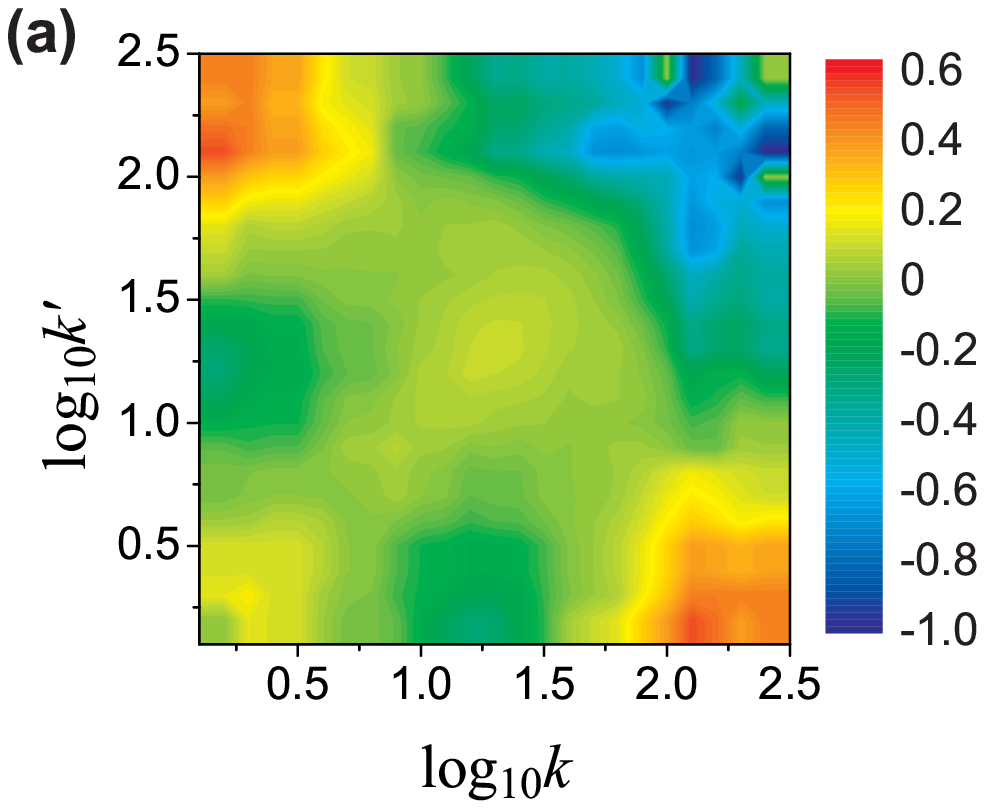}}
\end{minipage}\hfill
\begin{minipage}[!t]{0.5\linewidth}
\flushleft{\epsfxsize=6.3cm \epsfbox{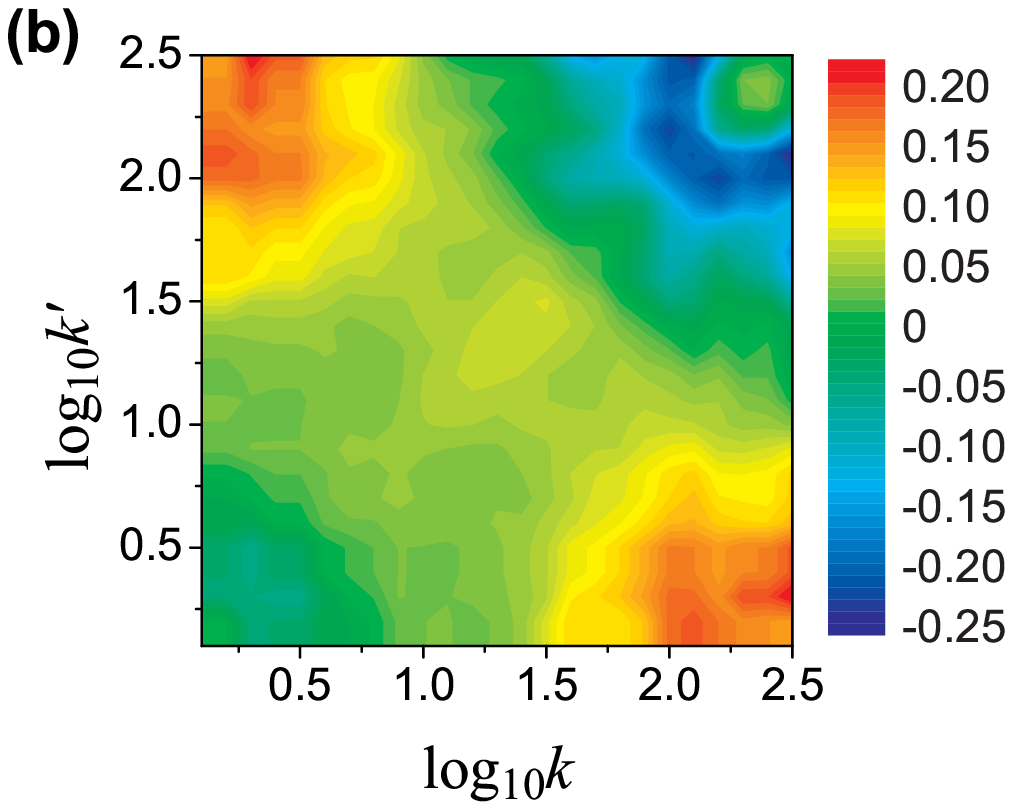}}
\end{minipage}\hfill
\caption{(a) Comparison between the degree correlation profiles of the 
yeast PIN and (b) the model network. The color code denotes the value
of $\log_{10}[P(k,k')/P_{\rm random}(k,k')]$. The randomized networks
are generated by the switching method \cite{maslov} 
that conserves the degree sequence.\\
}
\label{corr}
\end{figure*}

{\em Results}--- Now we compare the simulation results of our model. 
In typical simulations, 
we employed $\alpha=0.8$ and $\delta=0.7$. The value of $\delta$ was 
chosen to accommodate the fact that superfamilies exhibit extensive
sequence diversity~\cite{todd}. The value of $\alpha$ was set to match
the empirical value of the average degree of the PIN,
$\langle k\rangle\simeq 6.4$. Also, we matched approximately the numbers
of protein families and proteins with those of budding yeast, as we
described before.
The results obtained from the model show 
good agreements with the empirical data as shown in Fig.~2 and Table \ref{tab1}. 
In Fig.~2, we also show the results with the model without implementing
FC, which is similar to the model of Sol\'e et al.~\cite{sole}.
One can clearly see that without FC, we cannot 
account for the clustering and the degree correlation characteristics.
We also examine the full degree-correlation profile of 
the joint probability $P(k,k')$ that two proteins with degrees $k$ and
$k'$ are connected to each other. 
The degree-correlation intensity is quantified by $P(k,k')/P_{\rm random}(k,k')$,
the ratio with the joint probability in the randomized ensemble of
the original network \cite{maslov,sole03}.
As shown in Fig.~3, the profile obtained from the model 
has a pattern that is quite similar to that of the empirical yeast PIN.

\begin{figure}[t]
\centerline{\epsfxsize=\linewidth \epsfbox{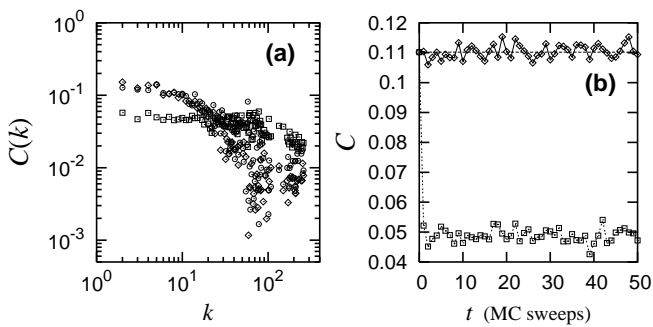}}
\caption{Network randomization test with and without FC.
(a) Clustering function $C(k)$ and 
(b) the clustering coefficient $C$ as functions of 
the number of edge shufflings are shown.
Symbols are for the unperturbed model network ($\bigcirc$), 
the network shuffled with FC ($\diamond$),
and the network shuffled without FC ($\Box$).
The horizontal line in (b) corresponds to the value of the clustering
coefficient in the unperturbed model network.
}
\end{figure}

To get further support for the relevance of the FC constraint,
we performed a network randomization test. We randomized the model network
by using the conventional edge switching method \cite{maslov}, but with the 
FC constraint. That is, when we are to switch the interactions
between the protein pairs, only the switching attempts that preserve 
FC are accepted. In this way, we can filter out the role of
FC. In Fig.~4, we show the results of randomization. We find that the
high clustering property of the network is preserved with randomization
with FC, but not without FC. Without FC, the clustering coefficient
drops as soon as we shuffle the network, as can be seen in Fig.~4(b). 
Thus, we conclude FC, indeed, plays a crucial role in PIN evolution.

\begin{figure}[t]
\centerline{\epsfxsize=9.5cm \epsfbox{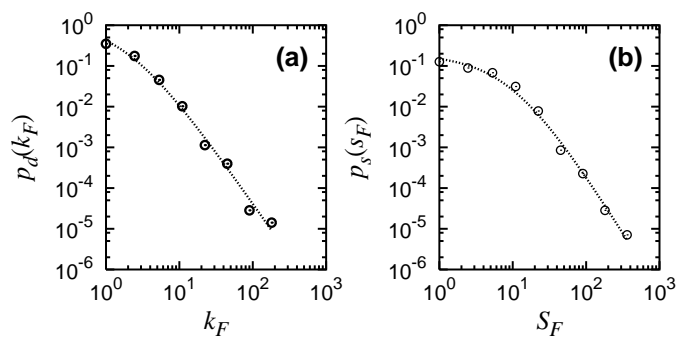}}
\caption{Simulation results for the protein family network: 
(a) The family degree distribution $p_d(k_F)$ and 
(b) the family size distribution $p_s(s_F)$.
The dotted lines in (a) and (b) are fit lines to Eq.~(\ref{pk}).
}
\end{figure}

Finally, we check the properties of the PFN. In Fig.~5, we show the 
degree distribution of the PFN and the family size distribution 
generated {\it in silico}. The degree distribution of the PFN follows 
a similar form to Eq.~(2), but with a different value of the exponent,
$\gamma_f\approx 3$. The family size distribution also follows a power 
law with an exponent of 3$\sim$4. 

In summary, we have introduced an {\em in-silico} model for PIN 
evolution. The model network is composed of the PIN and the PFN. 
In the early stage of evolution, the PIN and the PFN coevolve, 
and in the later stage, the PFN becomes fixed.
The evolution proceeds by the three major mechanisms
previously proposed, duplication, divergence, and mutation.
However, it is constrained by FC and 
follows a modified preferential attachment rule in the domain abundance,
which is the new feature of our model.
We have checked various structural properties of the model network, finding 
that they show good agreements with those of the integrated empirical data 
of the yeast PIN. 
Finally, it would be interesting to apply our model to higher eukaryotes,
as the data for the protein interactions are accumulating for the 
multicellular species such as the nematode worm {\em Caenorhabditis elegans} 
\cite{vidal} and the fruit fly {\em Drosophila melanogater} \cite{giot}.
\\

\begin{acknowledgments}
The authors would like to thank J.~Park for helpful conversation.
This work is supported by Korea Science and Engineering Foundation
grant No. R14-2002-059-01000-0 in the Advanced Basic Research Laboratory 
program and Ministry of Science and Technology grant No. M1 03B500000110.
\end{acknowledgments}

\end{document}